# Constraining the Limitations of NEATM-like Models: A Case Study with Near-Earth Asteroid (285263) 1998 QE2

Samuel A. Myers[1], Ellen S. Howell[1], Christopher Magri[2], Ronald J. Vervack, Jr.[3], Yanga R. Fernández[4], Sean E. Marshall[5], and Patrick A. Taylor[6]
[1] Lunar and Planetary Laboratory, University of Arizona, 1629 E. University Boulevard, Tucson, AZ 85721, USA; sammyers@lpl.arizona.edu
[2] University of Maine Farmington, 173 High Street, Farmington, ME 04938, USA
[3] Johns Hopkins Applied Physics Laboratory, 11100 John Hopkins Road, Laurel, MD 20723, USA
[4] University of Central Florida, 4111 Libra Drive, Orlando, FL 32816, USA
[5] Arecibo Observatory/University of Central Florida, HC-03 Box 53995, Arecibo, Puerto Rico 00612, USA
[6] National Radio Astronomy Observatory/Green Bank Observatory, 1180 Boxwood Estate Road, Charlottesville, VA 22903, USA
Received 2022 August 10; revised 2022 November 28; accepted 2022 December 1; published 2023 January 10

## Abstract

Near-Earth asteroids (NEAs) are a key test bed for investigations into planet formation, asteroid dynamics, and planetary defense initiatives. These studies rely on understanding NEA sizes, albedo distributions, and regolith properties. Simple thermal models are a commonly used method for determining these properties; however, they have inherent limitations owing to the simplifying assumptions they make about asteroid shapes and properties. With the recent collapse of the Arecibo Telescope and a decrease of direct size measurements, as well as future facilities such as LSST and NEO Surveyor coming online soon, these models will play an increasingly important role in our knowledge of the NEA population. Therefore, it is key to understand the limits of these models. In this work we constrain the limitations of simple thermal models by comparing model results to more complex thermophysical models, radar data, and other existing analyses. Furthermore, we present a method for placing tighter constraints on inferred NEA properties using simple thermal models. These comparisons and constraints are explored using the NEA (285263) 1998 QE2 as a case study. We analyze QE2 with a simple thermal model and data from both the NASA IRTF SpeX instrument and NEOWISE mission. We determine an albedo between 0.05 and 0.10 and thermal inertia between 0 and 425 J m$^{-2}$ s$^{-1/2}$ K$^{-1}$. We find that overall the simple thermal model is able to well constrain the properties of QE2; however, we find that model uncertainties can be influenced by topography, viewing geometry, and the wavelength range of data used.

*Unified Astronomy Thesaurus concepts:* Asteroids (72); Asteroid surfaces (2209); Near-Earth objects (1092)

## 1. Introduction

Asteroids were once derided by astronomers as the "vermin of the sky," but they now form an important piece of our efforts to understand our own solar system. Understanding their sizes, albedo distributions, and regolith properties is key for investigations into many aspects of solar system science, including solar system formation, main belt asteroid orbital evolution, surface processes on airless bodies, and understanding our meteorite collection. Near-Earth asteroids (NEAs), in particular, are excellent targets for these efforts owing to their proximity to Earth.

In addition to understanding the albedos and regoliths of these objects, accurately measuring the sizes of NEAs is pivotal for planetary defense initiatives—the area of study focused on preventing catastrophic asteroid impacts with Earth. This is because the size of an object is directly related to the energy of impact (Morrison & Teller 1995), which determines the impact severity. Thus, observation and modeling techniques that provide estimates of these properties are key for understanding the NEA population.

There are a few methods for obtaining size estimates and other physical properties from NEA observations. Radar images, detailed thermophysical models, and simple thermal models can all be used to obtain size estimates. All of these methods, along with light-curve measurements, can also place constraints on other physical properties of asteroids. Other methods, such as direct imaging (Dollfus 1971; Marchis et al. 2006; Marchis & Vega 2014), stellar occultations (Millis & Dunham 1989; Arai et al. 2020), and spacecraft encounters exist (Belton et al. 1992, 1996; Veverka et al. 2000; Lauretta et al. 2019) but are only applicable in rare cases. Of the more common methods, radar images can provide a size estimate without other information (Ostro 1985). Radar observations can also be used to construct detailed models of the asteroid's shape (Hudson & Ostro 1994; Magri et al. 2007, 2011; Nolan et al. 2013). Light-curve measurements can also produce shape models, although they are often less detailed than radar-derived shape models and do not include an absolute size scale (Ďurech et al. 2012 and references therein). These shape models can be coupled with thermal spectra to constrain other physical properties of the asteroid as well, such as thermal inertia or surface roughness (Marshall et al. 2017; Howell et al. 2018; Jones 2018; Hinkle et al. 2022).

Historically, the Arecibo Telescope has been a source of numerous NEA radar observations. The Arecibo Telescope detected over 900 NEAs and made size estimates of roughly 400 of those (Howell et al. 2020). However, with the recent loss of the Arecibo Telescope, there will be a lack of direct size and shape measurements of NEAs. (Although Goldstone is able to make radar measurements, it has a lower sensitivity and less availability for targets of opportunity.) As a result, in the future there will be a greater reliance on other methods to understand the physical properties of NEAs. These methods will necessarily be models, like simple thermal models, that assume asteroid shapes or use less well-constrained shape models.

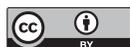

Simple thermal models, such as the Standard Thermal Model (Lebofsky et al. 1986; Lebofsky & Spencer 1989) and the Near-Earth Asteroid Thermophysical Model (NEATM; Harris 1998), are a convenient method for obtaining NEA sizes and physical properties in part because they are easy to run. They require only visible and thermal infrared data and are computationally fast. For this reason, they are already commonly used to analyze data collected by large survey missions like NEOWISE (Mainzer et al. 2011b) and ExploreNEOs (Trilling et al. 2010). Due to the large volume of data collected by these types of surveys and the sparse amount of data collected on any single object, simple thermal models are often the only practical way to quickly interpret the data. In these cases, simple thermal models are used to identify both scientifically interesting and potentially dangerous NEAs (e.g., Trilling et al. 2010).

However, simple thermal models make simplifying assumptions about the asteroid's shape and surface that can result in inaccuracies and thus poor constraints of inferred NEA properties. This is especially relevant for determinations of asteroid sizes—values that are pivotal for planetary defense activities. Simple thermal models can only make direct determinations of asteroid sizes in specific cases. If absolute photometry in both the visible and infrared is available, size can be solved for directly. However, these estimates require assuming that the visible and infrared data were acquired at similar viewing geometries. This assumption is often made with models employing NEOWISE or ExploreNEOs observations. Alternatively, if only normalized flux is available, then the size must be estimated from the modeled albedo in combination with the absolute magnitude, $H$. In this case, the estimates are subject to uncertainties in the magnitude (Bowell et al. 1989; Jurić et al. 2002; Vereš et al. 2015), as well as typically large error bars in the inferred albedo, producing poor constraints. In fact, recent work has shown that there are inconsistencies between sizes derived from NEOWISE data using these models and sizes derived using other methods (Howell et al. 2012; Taylor et al. 2014; Masiero et al. 2019; Taylor et al. 2019; Masiero et al. 2021).

In this paper, we seek to better understand the limitations of simple thermal models, such as NEATM, by comparing simple thermal model results to more complex thermophysical models, radar data, and other existing analyses of a given object. We also present a method for placing tighter constraints on inferred NEA properties using these simple thermal models. We use a simple, NEATM-like model (Section 3) to model the observed NEA, and the consistency of the best-fit parameters is then checked by comparing the models to normalized flux data collected across multiple nights that represent a range of viewing geometries. We also compare the models to the absolute photometry collected by the NEOWISE spacecraft. By observing an object across multiple viewing geometries and combining normalized flux spectra with absolute photometry, we are able to place tight bounds on modeled NEA properties. These simple thermal model results are then compared to model results from SHERMAN (Magri et al. 2018), a complex thermophysical model; radar measurements; and other observations and analyses of the given object. These comparisons allow us to place constraints on the overall limitations of the simple thermal model and identify key factors that influence uncertainties in simple thermal model results.

This analysis is performed on the well-studied NEA (285263) 1998 QE2 (hereafter referred to as QE2). QE2 is a spheroidal, binary NEA system, with an existing radar-derived shape model (Springmann et al. 2014). The secondary has a diameter ∼25% that of the primary (Springmann et al. 2014) and thus contributes only 6% of the total flux. Therefore, the primary object dominates the thermal emission from the system, and we neglect the secondary in our analysis. QE2 is an Xk-type asteroid in the Bus−DeMeo taxonomy, as derived from our SpeX prism spectra and a visible spectrum obtained by Hicks et al. (2013).

As part of our investigation into the limitations of the NEATM-like model, we find a discrepancy in the currently accepted $H$-magnitude for QE2. We find that the current value is inconsistent with the size derived from the radar measurements of QE2. We investigate this discrepancy and discuss implications. As part of this investigation, we compare our results to previous studies to understand QE2's composition and surface properties (Fieber-Beyer et al. 2020), as well as its spin state (Moskovitz et al. 2017). These comparisons allow us to further benchmark the uncertainties in the results of our method for placing tight constraints on NEA properties derived with simple thermal models.

In Section 2 we discuss the data used for our analysis. In Section 3 we describe our simple, NEATM-like model, and in Section 4 we present the results for QE2 from this model. In Section 5 we describe our analysis of the uncertainties in these model results. We compare our simple, NEATM-like model results to model results from SHERMAN, radar data of QE2, and the results of other previous studies. We then discuss implications for the limitations of simple thermal models. We conclude with a summary of our results in Section 6.

## 2. Spectral and Radar Data

### 2.1. IRTF Observations

The primary data used to constrain our models are normalized flux spectra obtained with the SpeX instrument at the NASA IRTF (Rayner et al. 2003). We use normalized flux, as it has smaller uncertainties relative to absolute photometry. These observations are carried out as part of our ongoing investigation into the physical properties of NEAs. We observed using both prism mode (0.8–2.5 μm) and Long-Wavelength Cross-Dispersed (LDX) 1.9 mode (2.2–4.1 μm). Note that the observations of QE2 presented here were done before the upgrade to SpeX that expanded the wavelength ranges of all settings.

For QE2, observations were carried out over six nights, from 2013 May 30 to 2013 July 10. Over this time, the solar phase angle of QE2 varied from 18°.0 to 39°.7, which let us observe different viewing geometries and illumination states. As a result, we see the thermal emission at different local times of day. This is important because it allows us to check the consistency of the fit parameters (Section 3). The various sub-Earth locations of QE2 that we observed are shown in Figure 1. A summary of the observational parameters for our six nights of SpeX data is shown in Table 1.

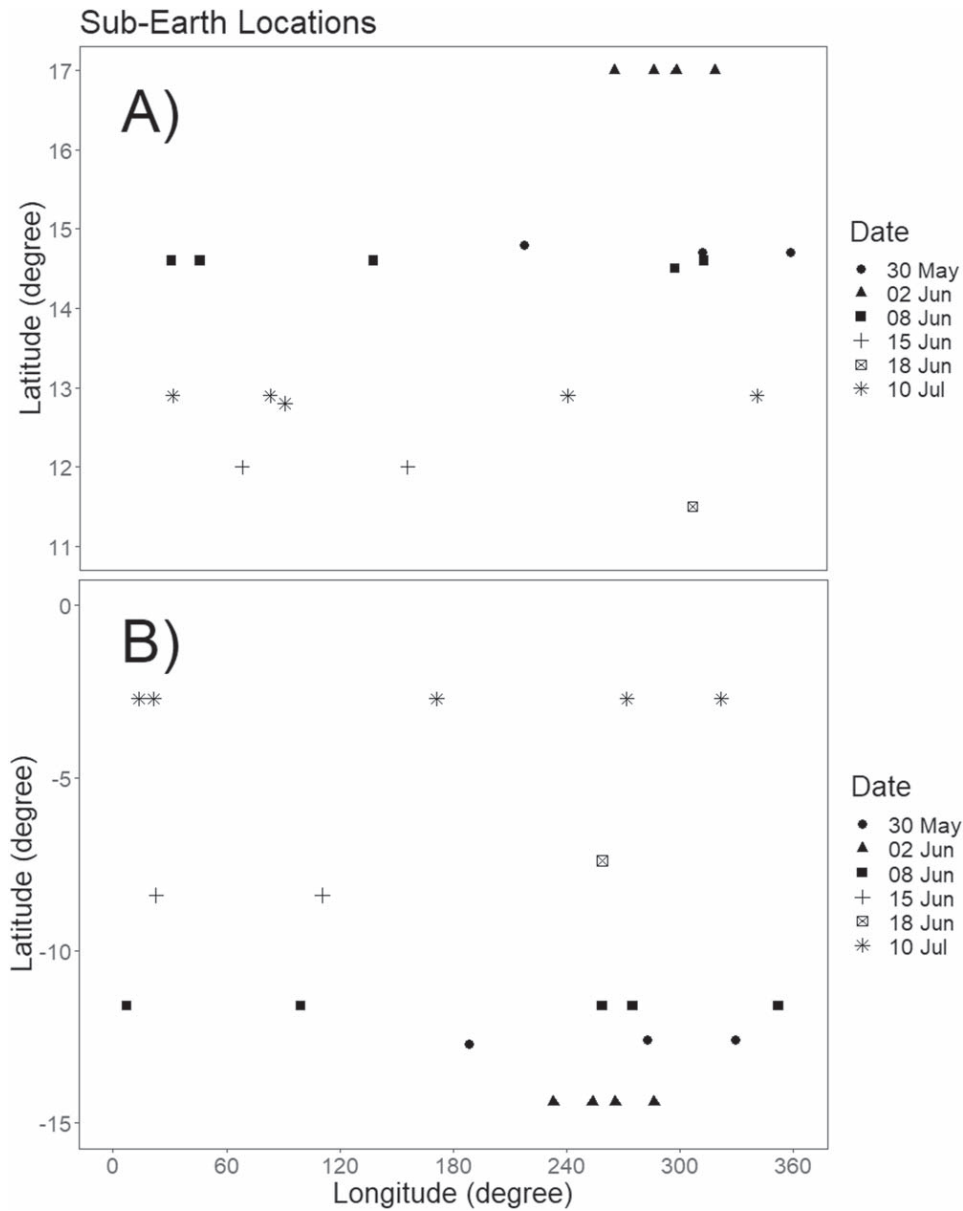

**Figure 1.** Sub-Earth locations on QE2 during observations as determined by a radar shape model (Springmann et al. 2014). (a) The pole solution with the "bumpy" topography in the northern hemisphere. (b) The pole solution with the topography partially in the southern hemisphere (Section 2.3). The range of sub-Earth locations observed indicates that QE2 was observed across multiple different viewing geometries. This range of observations is key for constraining QE2's parameters using our NEATM-like model.

All SpeX observations were done in pairs, nodding the telescope along a 15″ slit. We used exposure times of 15 s for our LXD data and 10–30 s for our prism data. The data were processed using the Spextool software package (Cushing et al. 2004), and the spectra were extracted from summed images. In addition to the object, we observed solar-analog stars in a similar manner. At least one was a nearby G star within ∼5° of the object on the sky. All stars were compared to a well-characterized solar analog star on each night, and their spectra were corrected for slight spectral slope variations if necessary. Each asteroid–star pair was combined in a ratio after correcting each for atmospheric absorption lines. The spectra were then determined using a weighted average over all asteroid–star pairs and binned to form the final spectra. Bad data points were flagged and excluded from the fitting and averaging process. The detailed methods for this entire process are given in Howell et al. (2018).

The data are broken up across each night into several independent sets of roughly 20–30 minutes each to sample different areas of the surface. QE2 has a rotation period of $4.749 \pm 0.002$ hr (Springmann et al. 2014), meaning that each spectrum is separated by roughly 25°–40° of longitude at the equator. The sub-Earth latitudes and longitudes at the midtimes of the observations are shown in Figure 1. These sub-Earth coordinates are calculated using the shape model of Springmann et al. (2014). The LXD data for each of the six nights are shown in Figure 2. Each spectrum is normalized at 1.6 μm to give normalized flux. (Note that there is no significant thermal

**Table 1**
Summary of Observations, Including Values Input Directly into the NEATM-like Model

| Date | Set | Midtime | $r_H$ (au) | $\Delta$ (au) | $\alpha$ (deg) | Instrument |
| --- | --- | --- | --- | --- | --- | --- |
| 2013 May 30 | A | 06:46:50 | 1.046 8 | 0.040 3 | 34.3 | SpeX |
| 2013 May 30 | B | 07:22:08 | 1.046 8 | 0.040 3 | 34.2 | SpeX |
| 2013 May 30 | C | 08:36:57 | 1.049 8 | 0.040 2 | 33.9 | SpeX |
| 2013 Jun 02 | A | 06:51:57 | 1.052 2 | 0.040 1 | 18.3 | SpeX |
| 2013 Jun 02 | B | 07:08:19 | 1.052 2 | 0.040 1 | 18.3 | SpeX |
| 2013 Jun 02 | C | 07:17:50 | 1.052 2 | 0.040 1 | 18.3 | SpeX |
| 2013 Jun 02 | D | 07:34:17 | 1.052 2 | 0.040 1 | 18.2 | SpeX |
| 2013 Jun 08 | A | 08:12:16 | 1.067 1 | 0.060 5 | 30.0 | SpeX |
| 2013 Jun 08 | B | 09:25:01 | 1.067 2 | 0.060 8 | 30.1 | SpeX |
| 2013 Jun 08 | C | 09:37:14 | 1.067 2 | 0.060 8 | 30.1 | SpeX |
| 2013 Jun 08 | D | 10:38:10 | 1.067 4 | 0.061 0 | 30.2 | SpeX |
| 2013 Jun 08 | E | 10:50:40 | 1.067 4 | 0.061 1 | 30.2 | SpeX |
| 2013 Jun 15 | A | 11:06:28 | 1.091 0 | 0.098 8 | 38.8 | SpeX |
| 2013 Jun 15 | B | 12:16:11 | 1.091 2 | 0.099 1 | 38.8 | SpeX |
| 2013 Jun 18 | A | 13:07:51 | 1.103 3 | 0.116 9 | 39.7 | SpeX |
| 2013 Jul 10 | A | 10:23:08 | 1.218 8 | 0.256 2 | 34.0 | SpeX |
| 2013 Jul 10 | B | 10:29:19 | 1.218 9 | 0.256 2 | 34.0 | SpeX |
| 2013 Jul 10 | C | 11:10:20 | 1.219 0 | 0.256 4 | 34.0 | SpeX |
| 2013 Jul 10 | D | 11:49:53 | 1.219 2 | 0.256 6 | 34.0 | SpeX |
| 2013 Jul 10 | E | 13:09:29 | 1.219 6 | 0.257 0 | 34.0 | SpeX |
| 2017 Jul 01 | A | 10:51:35 | 1.767 6 | 1.445 3 | 35.1 | NEOWISE |

**Note.** Set refers to different data sets on a given night. Midtime is the midtime of observation for the data set in UTC time. (Each SpeX observation spans roughly 20–30 minutes, while the NEOWISE observation spans 29 hr. Thus, each SpeX spectrum is separated by roughly 25°–40° of longitude.) $r_H$ is the Sun–object distance, $\Delta$ is the Earth–object distance, and $\alpha$ is the solar phase angle. Note that the observations are carried out across a range of solar phase angles and viewing geometries.

contamination at this wavelength.) We use normalized flux because the relative uncertainties are much smaller than for absolutely calibrated photometry. We cover the range from completely reflected to thermally dominated to ensure that our simple thermal model is well constrained in both regimes. This technique has the advantage of being more flexible but the disadvantage that the data are highly correlated in wavelength.

### 2.2. NEOWISE Observations

In addition to our SpeX data, we fit our simple thermal model to data collected by NEOWISE. Unlike the SpeX data, which measure normalized flux, NEOWISE measures absolute photometry. Thus, fitting our simple thermal model to the NEOWISE data allows us to check that the best-fit parameters are consistent with both the spectrum shape and calibrated flux values. This provides an additional independent check on the consistency of the simple thermal model and allows us to identify any potential issues with the model not observed when fitting normalized flux data alone.

We retrieve the NEOWISE data and associated uncertainties from the NASA/IPAC Infrared Science Archive (Mainzer et al. 2011a, 2014).[7] We do not use the raw images, but instead retrieve processed data that list the magnitudes and uncertainties for channels W1 (effective wavelength 3.4 μm) and W2 (effective wavelength 4.6 μm) for each time the object was observed. We remove data points that are flagged for potential contamination, such as by cosmic-ray hits, and average together all remaining observations. The uncertainty in the NEOWISE data is dominated by systematic errors and not statistical noise. All observations, except one, have similar uncertainties. We thus take a weighted average of the observations and adopt the variance of the overall data set, divided by the square root of the number of observations minus one, as our 1σ uncertainties. For QE2, all observations were taken over a short time interval such that the change in QE2's orbital position was minimal. Therefore, we averaged together all available observations, resulting in one averaged set of data points from eight individual observations that span roughly 29 hr and approximately six rotation periods. The individual observations are evenly distributed across the rotation phase. A summary of the observational parameters for the averaged observation is given in Table 1. A list of the individual observations is given in Table 2.

After retrieval, the data are then converted from NEOWISE magnitudes to $F_\lambda$ units following the procedures outlined in the WISE Data Processing Handbook (Wright et al. 2010; Cutri et al. 2012). For this process we apply a final blackbody color correction corresponding to a 221 K object. This blackbody temperature is determined by fitting ideal blackbody curves to the NEOWISE data in an iterative process until the corrected NEOWISE data and ideal blackbody curves converge. The blackbody temperature used for the initial correction is calculated using the theoretical blackbody temperature relation

$$\sigma_{\rm sb} T^4 = \frac{L_\odot (1-A)}{16 \pi r_H^2}, \quad (1)$$

where $L_\odot$ is the solar luminosity, $A$ is the Bond albedo, $r_H$ is the object–Sun distance, and $\sigma_{\rm sb}$ is the Stefan–Boltzmann constant.

---
[7] https://www.ipac.caltech.edu/doi/irsa/10.26131/IRSA144

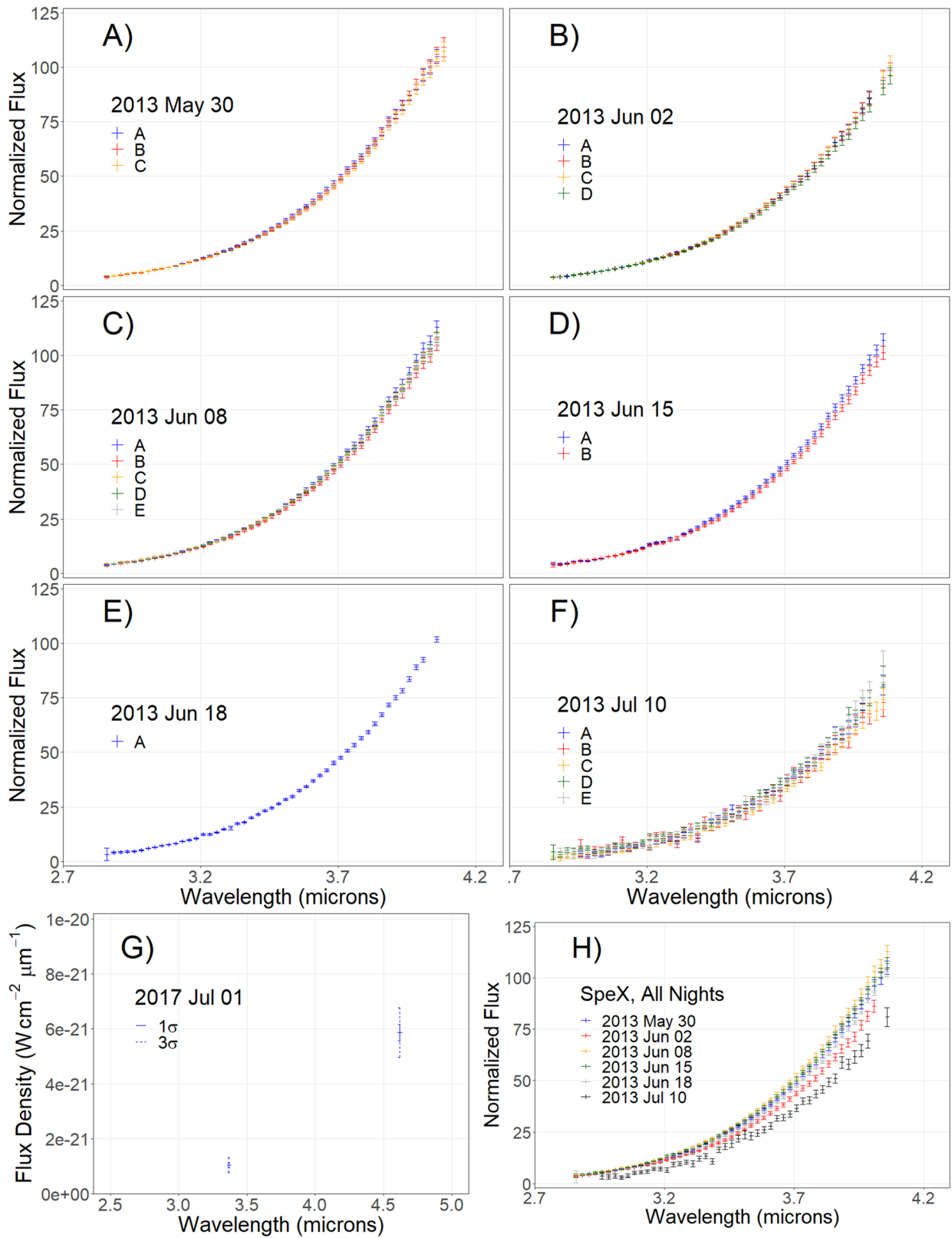

**Figure 2.** Processed LXD data sets for each night of observations with SpeX and NEOWISE. (a–f) SpeX data for each of the six nights. The different letters within each panel indicate different data sets collected each night (Table 1). The y-axis is normalized flux, normalized to 1.6 $\mu$m. (Note that there is no significant thermal contamination at this wavelength.) (g) NEOWISE data in absolute flux density. Note that the NEOWISE data are plotted over a different wavelength range. We plot both the 1$\sigma$ and 3$\sigma$ uncertainties. (h) The "A" data set for each night of SpeX data. These spectra highlight how different viewing geometries across the different nights produce a range of spectral slopes. We see that changes in viewing geometry produce changes in the spectra shape both within nights and across all nights of observations. Modeling these differences allows us to place tighter constraints on NEA properties.

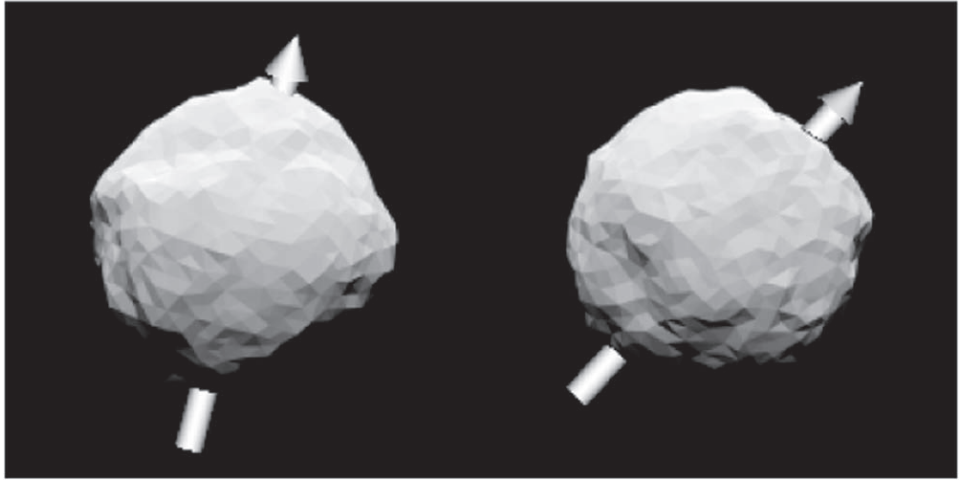

**Figure 3.** Sky views of QE2 on 2013 July 10 that show the radar shape model from Springmann et al. (2014). The arrows indicate the pole and spin direction. Left: the A solution with a pole position of $\lambda = 119°$ and $\beta = 55°$. Right: the B solution with a pole position of $\lambda = 158°$ and $\beta = 41°$.

**Table 2**
List of Individual NEOWISE Observations Used to Obtain the Single Averaged NEOWISE Data Set

| Date | Midtime | $m_1$ (mag) | $\sigma_1$ (mag) | $m_2$ (mag) | $\sigma_2$ (mag) |
|---|---|---|---|---|---|
| 2017 Jun 30 | 19:32:22 | 16.768 | 0.468 | 13.609 | 0.136 |
| 2017 Jun 30 | 22:41:03 | 16.256 | 0.253 | 13.844 | 0.158 |
| 2017 Jul 01 | 03:23:49 | 16.625 | 0.392 | 13.944 | 0.172 |
| 2017 Jul 01 | 06:32:19 | 16.966 | 0.535 | 13.658 | 0.131 |
| 2017 Jul 01 | 15:58:00 | 16.449 | 0.338 | 14.193 | 0.292 |
| 2017 Jul 01 | 19:06:30 | 16.927 | 0.522 | 13.694 | 0.135 |
| 2017 Jul 01 | 22:15:00 | 16.557 | 0.476 | 13.801 | 0.161 |
| 2017 Jul 02 | 01:23:41 | 16.539 | 0.375 | 13.770 | 0.124 |
| 2017 Jul 01 | 10:51:35 | 16.528 | 0.092 | 13.758 | 0.051 |

**Note.** Midtime is the midtime of observation in UTC time. $m_1$ and $m_2$ are the NEOWISE reported magnitudes for W1 (effective wavelength 3.4 $\mu$m) and W2 (effective wavelength 4.6 $\mu$m), respectively. $\sigma_1$ and $\sigma_2$ are the NEOWISE reported magnitude uncertainties for W1 and W2, respectively. The last row is the averaged observation.

The Bond albedo is estimated according to the method described in Lebofsky & Spencer (1989):

$$A = (0.29 + 0.684G)p, \quad (2)$$

where $G$ is the slope parameter in the HG magnitude system (Bowell et al. 1989) and $p$ is the visual geometric albedo. The standard assumption of $G = 0.15$ is used, and $p$ is taken from the model fits to the SpeX data. Note that since the fitting process is iterative, choices of the initial guess parameters do not strongly affect the final result. The end products of this conversion process are flux densities reported in units of W cm$^{-2}$ $\mu$m$^{-1}$, which match the units of our simple thermal model output. The final NEOWISE data for QE2 are shown in Figure 2. We show the data with both 1$\sigma$ and 3$\sigma$ uncertainties.

*2.3. Radar Shape Model*

As part of our investigation into the limitations of simple thermal models, we compare the results of our NEATM-like models to many other data sources and models, including radar images and a radar shape model. The radar image is a direct measurement of the size that only depends on the viewing geometry and the speed of light. A spheroidal object, such as QE2, shows a radius in radar range at nearly all aspects and is a robust size estimate. We compare the radar size to sizes derived from our NEATM-like model, based on the magnitude and albedo. We emphasize that this information is not used as an input of our NEATM-like model and is only used to compare with our NEATM-like model results.

The radar shape model for QE2 is described by Springmann et al. (2014). The model is constructed using observations from the Arecibo Observatory and Goldstone. Data used were collected between 2013 May 31 and June 9, during QE2's close approach to Earth. These radar images are used to derive a shape model as described in Magri et al. (2011). A nonlinear iterative process is used to adjust synthetic radar images to match the observations by minimizing the difference between them. This process is described in detail in several papers for other objects (Magri et al. 2011; Nolan et al. 2013). The shape model of QE2 is preliminary, and the complete analysis is beyond the scope of this paper. However, the derived diameter of the principal axes of QE2 is robust and reliable as a comparison to values obtained here. This analysis gives a diameter for QE2 of $3.2 \pm 0.3$ km and a diameter of the secondary of $800 \pm 80$ m. QE2 is spheroidal, with a few dominant surface features.

Springmann et al. (2014) find a rotation rate of $4.749 \pm 0.002$ hr for QE2 and two possible pole solutions, both of which are prograde. One of these solutions, which we refer to as the A solution, places most of the "bumpy" topography of QE2 in the northern hemisphere. This solution has a pole position of $\lambda = 119°$ and $\beta = 55°$, where $\lambda$ is the ecliptic pole longitude and $\beta$ is the ecliptic pole latitude. The second solution, which we refer to as the B solution, places the "bumpy" topography partially in the southern hemisphere. This solution has a pole position of $\lambda = 158°$ and $\beta = 41°$. Both solutions are shown in Figure 3.

### 3. NEATM-like Model

The simple thermal model we use to fit the data is based on the Standard Thermal Model (Lebofsky et al. 1986; Lebofsky & Spencer 1989) and NEATM (Harris 1998). Our model is a variation of these models that we call our NEATM-like model (Howell et al. 2018). Like these models, for a given set of asteroid parameters, our NEATM-like model produces a theoretical thermal emission spectrum of the object that can be fit to any subset of the visible to near-IR spectra of an asteroid. However, our model also utilizes a simple incorporation of the rotation rate of the object that allows us to model the thermal inertia. The thermal inertia is a measurement of how well the object's surface retains heat energy from the Sun and is measured in J m$^{-2}$ s$^{-1/2}$ K$^{-1}$ (hereafter referred to as TIU for thermal inertia units). By determining the thermal inertia, in combination with the rotation rate, our NEATM-like model is able to account for differences across the day and night sides of an object. Thus, when incorporating many different observations of a single object, taken at different viewing geometries, we are able to model how changes in thermal inertia affect the thermal emission of an object. Overall, this incorporation allows us to get a more robust picture of the properties of the object. We note that other than this addition, this model is functionally similar to the standard NEATM model.

In addition to incorporating these parameters, our model also makes the typical assumption of a spherical shape for the asteroid. It also assumes subsolar and subobserver points on the asteroid's equator and prograde rotation at a fixed rotation rate. (The NEATM-like model does not account for shape effects, and the radar-derived shape model of Springmann et al. 2014 is only used to compare to the NEATM-like model results to investigate the limitations of the NEATM-like model.) The model also incorporates a free-floating beaming parameter—a scaling factor between the observed and predicted flux from the asteroid. This factor accounts for additional effects not included in the model, such as surface roughness, deviations from a spherical shape, local shadowing, and nonzero obliquity. The beaming parameter generally ranges between ∼0.5 and 2.0, with higher values usually occurring at higher phase angles or for more irregularly shaped asteroids.

Overall, our model includes three free-floating parameters: the visual geometric albedo, thermal inertia, and beaming parameter. The output of each run is a model spectrum of the asteroid, based on the input parameters, for each combination of the free-floating parameters. Thus, identifying best-fit parameters requires inspecting the model results and making direct comparisons to the data.

For a given object, the consistency of these fit parameters can be checked by comparing the results to thermal infrared data collected across multiple nights that represent a range of viewing geometries. This is important because many combinations of albedo, thermal inertia, and beaming parameter can fit any individual observation. By comparing model results for a single object to data taken at multiple different viewing geometries of that object, we can thus identify consistent values of albedo and thermal inertia that fit every observation, breaking degeneracies in the solution. The beaming parameter is allowed to vary, as it is expected to change in value across each observation. Thus, across multiple different viewing geometries, only a tight range of albedo and thermal inertia values will fit every observation. This is true even when the beaming parameter is allowed to vary, as more extreme deviations in albedo or thermal inertia would require increasingly extreme values of the beaming parameter to fit the observations, and realistic beaming parameters are generally constrained to the range of ∼0.5–2.0 (Delbó et al. 2003). Note that these comparisons are done solely to constrain the parameter fits of the NEATM-like model and are separate from the comparisons done as part of our investigation into the limitations of the NEATM-like model (Section 5).

The fixed model inputs for our NEATM-like model are the object's rotation period, a visible-to-near-IR reflectance ratio, Earth–object and Sun–object distances, solar phase angle, emissivity, and spherical equivalent diameter. For QE2, we use a rotation period of $4.749 \pm 0.002$ hr that was used by a previously derived radar shape model (Springmann et al. 2014). We also use a spherical equivalent diameter of 3.2 km from the same shape model. We note that since the shape of QE2 is very close to spherical, the assumption of spherical shape by the NEATM-like model is a very good assumption. The visible-to-near-IR reflectance ratio is estimated to be 1.127 using our SpeX prism spectra and a visible spectrum obtained by Hicks et al. (2013). This is a color correction factor used to relate the visible albedo to the near-infrared albedo at 1.6 $\mu$m, chosen as the normalization wavelength of the spectra. Earth–object and Sun–object distances, as well as solar phase angle, are calculated for each observation using JPL Horizons[8] based on the midtime of observation for each data set. These values are listed in Table 1. The emissivity is set to 0.9.

## 4. NEATM-like Model Results

We generate NEATM-like models for each of our normalized flux SpeX data sets and our single absolute photometry NEOWISE data set. Models are generated across a wide range of albedos, thermal inertias, and beaming parameters. Models are then compared to the data using an objective function to constrain QE2's properties. For any given data set, models of varying parameters change monotonically (Figure 4). These models are sorted by calculating a reduced $\chi^2$ between the model and the data. When performing this calculation, we only consider data points between 3.00 and 4.05 $\mu$m, as this is the region of strongest thermal emission without significant overlap with atmospheric water vapor lines. For the NEOWISE data set, both NEOWISE data points are used.

It is important to note that the reduced $\chi^2$ value we calculate is not a formal $\chi^2$, as it does not reach a minimum at unity and does not go up by a value of 1 when the model is 1$\sigma$ away from the data. This is because the uncertainties in the data are dominated by systematic effects, not statistical errors. The data points are not independent, as they are strongly correlated in wavelength and are affected by changing effects such as atmospheric conditions on different days, viewing geometry, and rotational changes of the asteroid. As a result, this calculation can be used to sort the goodness of fit of models for a given data set but cannot be used to compare models across data sets. Thus, for each data set, we use this method to identify the range of albedos and thermal inertias that produce models that lie within the 1$\sigma$ uncertainties of the data. Figure 5 shows the variation in models that were accepted to fit the data for each data set. (Note that for the NEOWISE data we also examine the models that fit within the 3$\sigma$ uncertainties. This range is also shown for the NEOWISE data.) Any models within the shown region are considered to fit the data. All other models for the given data set are discarded, as they are poor fits to the data.

---

[8] https://ssd.jpl.nasa.gov/horizons/

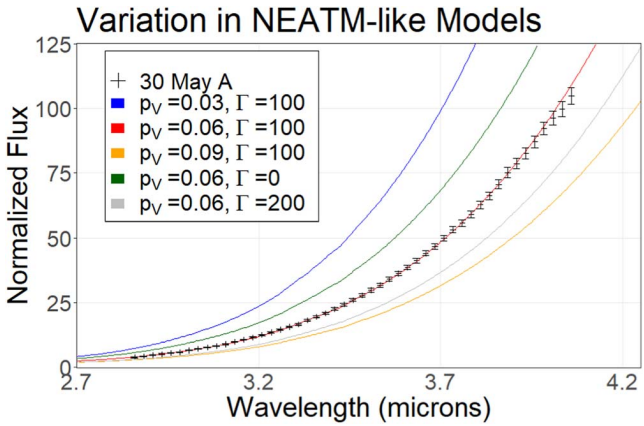

**Figure 4.** A range of NEATM-like models compared to one of our SpeX data sets. As either the albedo or thermal inertia changes monotonically, the models correspondingly change monotonically across the data. This property allows us to identify a range of models that fit the data and is typical to all of our data sets. All models that fall within the 1σ error bars of the data would be considered good fits to the data. As such, in this case only the $p_V = 0.06$ and $\Gamma = 100$ TIU model would be considered a good fit. Models shown here all have $\eta = 0.86$. Changes in beaming parameter can also monotonically affect how the models fit to the data.

For each data set, we then have a range of albedos and thermal inertias that can be said to fit that given data set. These individual fit spaces are shown in Figure 6. (For the NEOWISE data, we show the models that fit the 1σ uncertainties.) Overall, we have 21 such data sets: 20 data sets spread across six nights of IRTF SpeX observations, and 1 set of NEOWISE data. To identify the range of albedos and thermal inertias that describe QE2 overall, we then search for the region of overlap between each of these 21 different model sets. These results are shown in Figure 7. There is a clear section in the parameter space of models that fit nearly every data set. We define this region as the best-fit space.

All models within this space are consistent with the SpeX data, but do not fit the NEOWISE data with 1σ uncertainties. We then examine models that fit the NEOWISE data with 3σ uncertainties, and find that all models within the best-fit space are consistent with the NEOWISE data. This could be because the NEOWISE observations were taken at a much higher Sun–object distance than the SpeX data. As a result, QE2 was much colder at the time of these observations which may be introducing complexities to the thermal emission that our simple thermal model is not able to capture. Such effects may be better understood using a more complex thermophysical model, however a full investigation of this effect is beyond the scope of this work.

Overall, our analysis gives best-fit ranges of 0.05–0.10 for the visual geometric albedo and 0–425 TIU for the thermal inertia. Note that there is a correlation such that higher thermal inertias require lower albedos. Results are summarized in Table 3.

In general, we find a preference for lower beaming parameters of ~0.55–0.80. Beaming parameter results are shown in Figure 8. We remind the reader that we expect the beaming parameter to change across observations, and so we do not attempt to fit for a single overall value of the beaming parameter. These values are calculated by taking the best-fit beaming parameter value for a fixed albedo of 0.07 and a fixed thermal inertia of 150 TIU. These values are chosen because they are near the center of the best-fit region. The NEOWISE beaming parameters are calculated using the 3σ uncertainties as they are the results consistent with the SpeX data. As expected, the beaming parameter is generally higher for higher phase angles. The exceptions to this trend are July 10 and the NEOWISE data, both of which have substantially greater $r_H$ and Δ values than the other nights. These larger distances also explain the noisier data observed on July 10.

## 5. Limits of the NEATM-like Model

In calculating our best-fit model ranges, we compared our model results across many data sets taken at different viewing geometries of QE2 (Figure 1). These comparisons have allowed us to place tighter constraints on our modeled albedo and thermal inertia than would be possible with single observations. These albedos and thermal inertias can then be compared to results from more complex thermophysical models, radar data, and other observations to identify how accurately the NEATM-like model was able to constrain the properties of QE2. Our model results also provide us with a range of beaming parameter values that change as a function of viewing geometry. Analyzing these changes in beaming parameter can allow us to identify the unmodeled factors limiting the accuracy of our NEATM-like model. Overall, by comparing our model results to previous studies of QE2, we can gain insight into the limitations of simple thermal models as applied to a single object. In the subsections below we walk through comparisons of our simple thermal model results to various other models and data sets. For each comparison, we discuss in what ways our simple thermal model results differ and discuss implications for the factors affecting the uncertainties of simple thermal model results.

### 5.1. Albedo, Size, and H-magnitude

Our modeled visual geometric albedo for QE2 of 0.05–0.10 is higher than but overlaps with previously published values of $0.03^{+0.03}_{-0.02}$ (Moskovitz et al. 2017) and $0.04 \pm 0.01$ (Fieber-Beyer et al. 2020). We can use our modeled albedo, in combination with a radar-derived size, to estimate QE2's H-magnitude. This is given by the relationship

$$H = -5 \log_{10}\left(\sqrt{p}\, \frac{D}{1329 \text{ km}}\right),  \quad (3)$$

where $p$ is the albedo and $D$ is the object diameter in kilometers (Pravec & Harris 2007, Equation (3)). Using the diameter of $3.2 \pm 0.3$ km given by Springmann et al. (2014) and our modeled albedo range of 0.05–0.10, we get an H-magnitude of 15.4–16.6. This value is lower than (but partially overlaps with) previously given H-magnitude values of 16.4 (Trilling et al. 2010) and 17.3 (Moskovitz et al. 2017) for QE2.

However, the radar shape model constrains the diameter with high accuracy. The radar-derived shape can be considered a true constraint on QE2's size, as size can be measured directly from a radar image (Ostro 1985). Figure 9 shows a radar image of QE2 taken by the Arecibo Telescope on 2013 June 10. The vertical extent of the image shows distance from the observer to the terminator of the object. Thus, the resolution of the pixels, combined with knowledge of the speed of light, directly gives the object's radius. In this image, QE2 covers 210 pixels in the vertical extent at 7.5 m pixel$^{-1}$, giving an apparent radius of 1575 m or a diameter of 3.15 km. However, using an H-magnitude of 17.3 and albedos of 0.05–0.10 gives a diameter

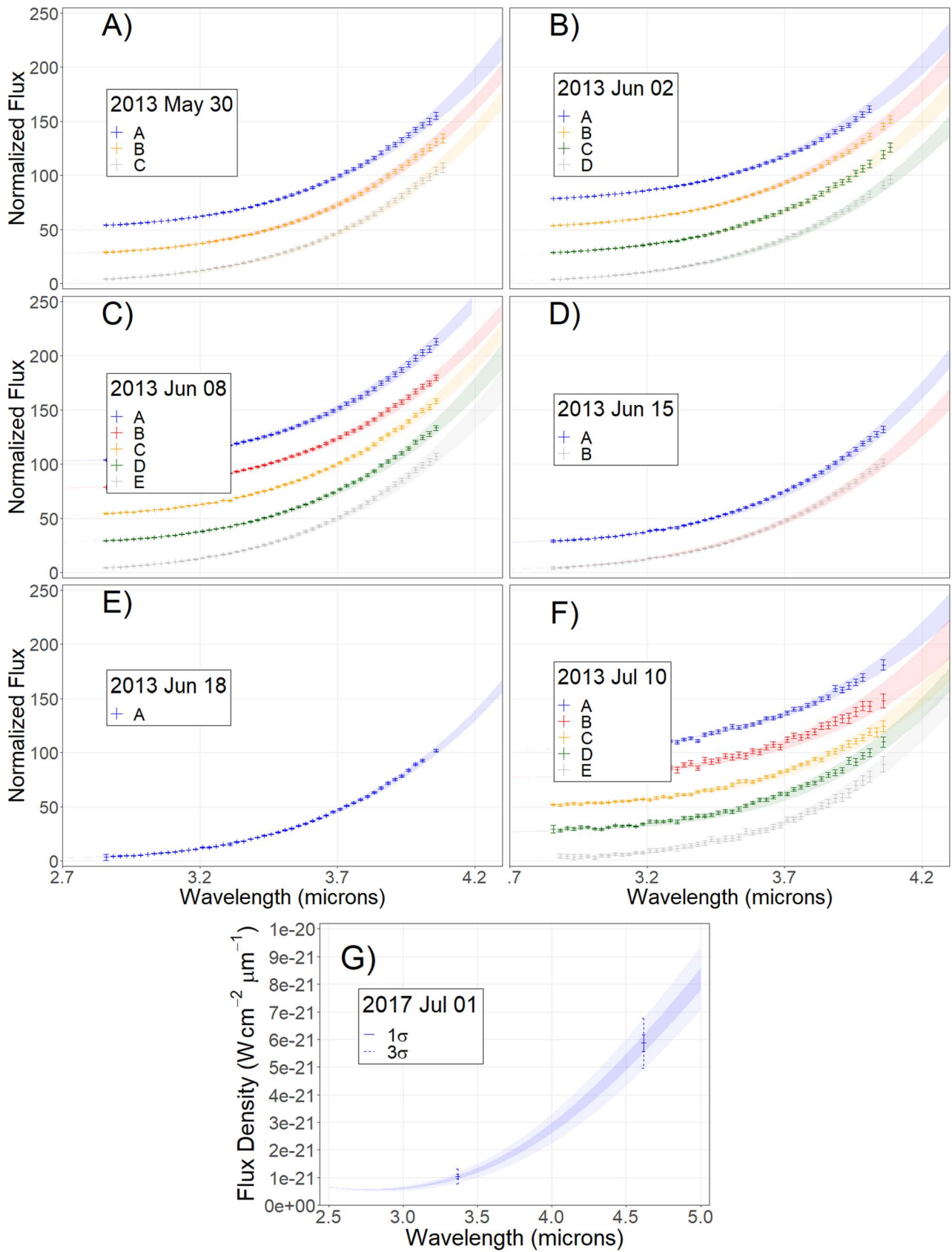

**Figure 5.** The variation in NEATM-like models that were accepted to fit the data for each data set. Any models within the shaded region are considered to fit the data. All other models for the given data set are discarded, as they are poor fits to the data. An objective function is used to identify which models fall within the shown region (Section 4). (a–f) SpeX data. The y-axis is normalized flux. The spectra are offset for clarity. (g) NEOWISE data in absolute flux density. Note that the NEOWISE data are plotted over a different wavelength range. For the NEOWISE data we examine models that fit within both the 1σ and 3σ uncertainties. Both regions are shown.

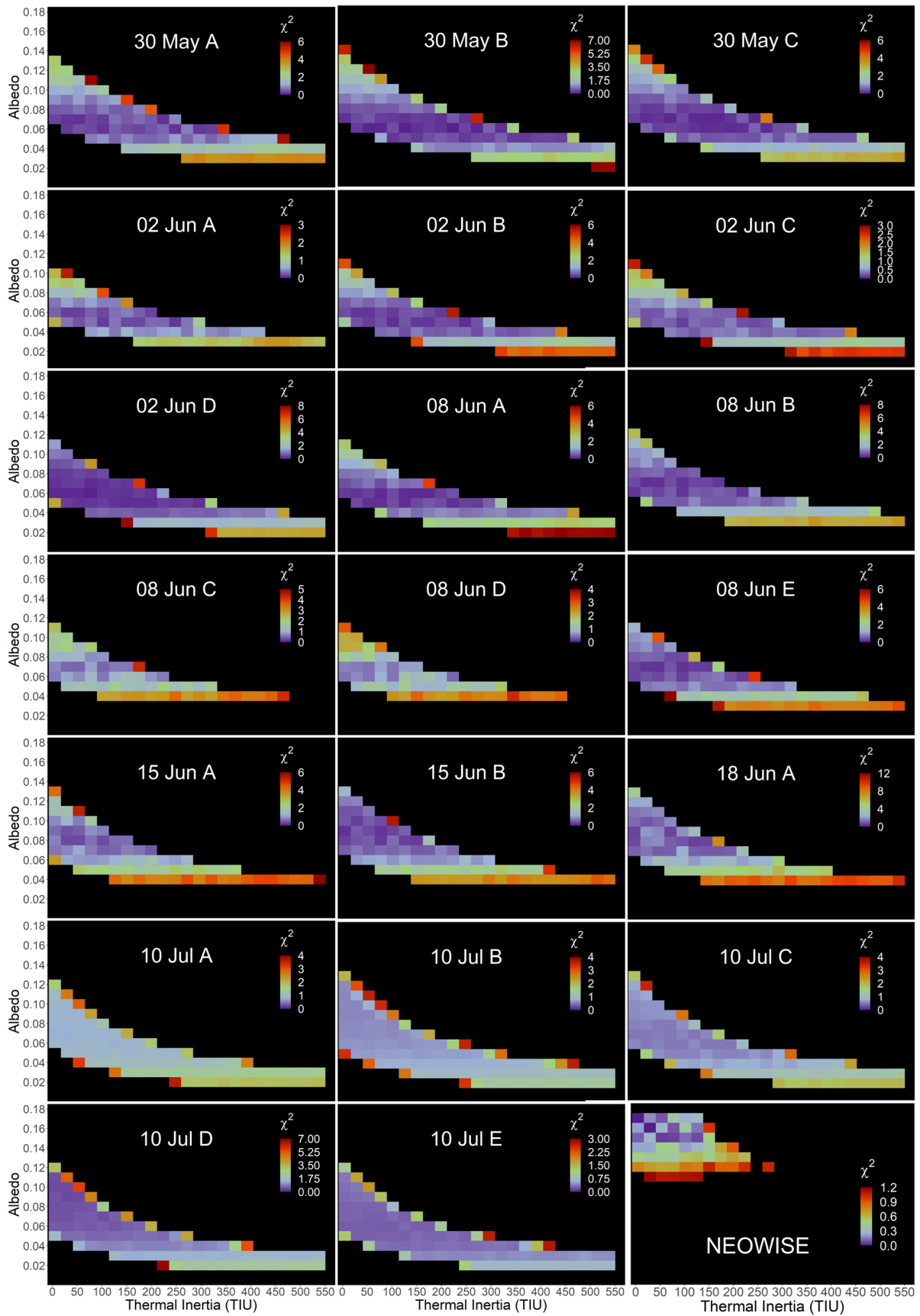

**Figure 6.** Reduced $\chi^2$ maps for each of the spectra as fit by our simple, NEATM-like model. Warmer colors mean higher values (worse fits), and cooler colors mean lower values (better fits). Note that different max values are used for different spectra, as the reduced $\chi^2$ are not directly comparable across different spectra (Section 4). Each $\chi^2$ map is equivalent to showing the range of models that fit a given data set. The fit space of the NEOWISE data corresponds to the 1σ uncertainties

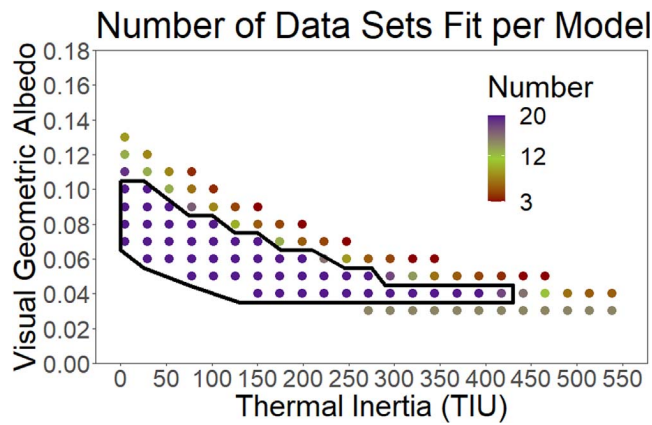

**Figure 7.** Final best-fit space for the visible geometric albedo and thermal inertia for QE2 using our simple, NEATM-like model. The color of the points represents the number of data sets that are fit by the associated parameter values. A cooler color means that the given parameters are consistent with more data sets. White indicates models consistent with ≤ 2 data sets. The black line outlines the region of best fit. This region corresponds to the region of overlap between all the individual model ranges found to fit each individual data set (Section 4). There is a correlation such that higher thermal inertias require lower albedos. All models were run with the same fixed model inputs listed in Section 3 and using ephemeris inputs listed in Table 1. This figure is generated using the results from the 1σ uncertainties on the NEOWISE data.

**Table 3**
Best-fit Model Ranges for the Three Free-floating Parameters of Our NEATM-like Model

| Parameter | Range |
| --- | --- |
| Albedo | 0.05–0.10 |
| Thermal inertia | 0–425 TIU |
| Beaming parameter | ∼0.55–0.80 |

**Note.** Albedo is visual geometric albedo. We expect the albedo and thermal inertia to be consistent across all data sets, and thus the ranges given represent the uncertainty in our model results. However, we expect the range of acceptable beaming parameters to change across observations, and thus the range given represents the range of values observed across all data sets.

between 1.5 and 2.1 km, well outside of the 1σ errors of the radar measurement.

We investigate this unusually large discrepancy in the $H$-magnitude by looking at existing observations. Using an $H$-magnitude value and an assumed $G$ value, we can calculate predicted apparent magnitudes. These predicted apparent magnitudes can then be compared to observed apparent magnitudes reported to the Minor Planet Center (MPC).[9] Ephemeris values are calculated for QE2 using JPL Horizons[10] at 1-day intervals throughout 2013. We then calculate predicted apparent magnitudes for the $H$-magnitude consistent with the radar-determined size and our modeled albedo, the $H$-magnitude used by Moskovitz et al. (2017), and a range of $G$ values from 0 to 0.15. This was done following the procedure in Bowell et al. (1989). These predicted apparent magnitudes are then compared to all the apparent magnitudes listed in the MPC. The results are shown in Figure 10. We see that $H$-magnitudes of neither 16.0 nor 17.3 perfectly match the data, but instead provide an upper and lower bound, respectively.

---

[9] https://minorplanetcenter.net/
[10] https://ssd.jpl.nasa.gov/horizons/

However, we note that an $H$-magnitude of 16.0 appears to provide a more reasonable fit than an $H$-magnitude of 17.3.

So what could be causing these $H$-magnitude differences? One possible explanation is related to the $G$ parameter. The $G$ parameter is often assumed to be 0.15 and is not fitted directly. Figure 10 shows that for $H = 16.0$ lower $G$ values fit better, while for $H = 17.3$ higher $G$ values fit better. For QE2, we would expect a lower $G$ value, as lower $G$ values are generally preferred for low-albedo objects owing to the smaller opposition effect. However, we note that the differences do not exceed ∼0.5 mag and thus cannot fully explain the discrepancy.

Another possible explanation is related to color effects; however, the color of QE2 is very close to solar, and thus this is also unlikely to be a large factor in this case. The discrepancy could also be due to the secondary contributing to the magnitude. Using the radar shape model (Springmann et al. 2014), we can calculate the effective diameter of the combined primary and secondary to be 3.3 ± 0.3 km. Using our modeled albedo range, this gives an $H$-magnitude difference of only ∼0.1 and thus is an ignorable contribution to the uncertainty. Therefore, none of these effects by themselves can fully explain the observed differences. Overall, given the accuracy of the radar measurement of the diameter and our tightly constrained albedo range, the true $H$-magnitude cannot be as high as 17.3. It is therefore likely that a better estimate of the $H$-magnitude lies somewhere in between 16.0 and 17.3. Furthermore, our analysis shows that higher H-magnitudes and thus lower albedos are likely favored for QE2, potentially further constraining the results from our simple thermal model.

### 5.2. Wavelength Range of Observations

We can also analyze our results by leveraging the large wavelength range of our observations. Our observations span 0.8–4.1 μm, and thus we are able to observe both the thermally dominated region of the spectra (≳3.0 μm) and the thermal tail (∼2.0–2.5 μm). We are therefore able to compare our model fits to both regions. This is notable because many studies (e.g., Moskovitz et al. 2017) rely only on the tail region. We show this comparison for a selection of our data sets in Figure 11.

We find that in nearly all cases the models that best fit the thermally dominated region also fit the tail region. However, for some dates (such as some data sets for 2013 July 10), an albedo increase of ∼0.02 relative to the model that fits the thermally dominated region is required to fit the tail region. This implies that QE2 may have an inhomogeneous surface and that we may be observing local thermal variations. Such variations could impart a wavelength-dependent change in the flux, thus creating the observed discrepancy. Another possibility is that some other effect, such as surface roughness, that our NEATM-like model does not account for may be causing this mismatch. This result is important because it shows the dangers of relying on only a limited spectral region to derive surface properties such as albedo.

### 5.3. Surface Topography

The potential effects of a surface inhomogeneity can be investigated by comparing our NEATM-like model results to results from a more complex thermal model. In addition to our NEATM-like models, we generate models using SHERMAN. SHERMAN is a more complex thermophysical model that takes account of the object's shape and that can separate the effects of obliquity and self-shadowing. (For a full description of SHERMAN, see Magri et al. 2018.) We give SHERMAN the radar-derived shape model of QE2 (Springmann et al. 2014), as well as our SpeX thermal infrared data. We also input a reflectance spectrum from our prism data, as well as a Hapke scattering function. SHERMAN has three free-floating parameters: visual geometric albedo, thermal inertia, and crater fraction. The crater fraction is a proxy for surface roughness and describes the fraction of each model facet covered with hemispherical craters, following the method of Lagerros (1998). SHERMAN outputs a modeled thermal spectrum that we then compare with our thermal infrared data.

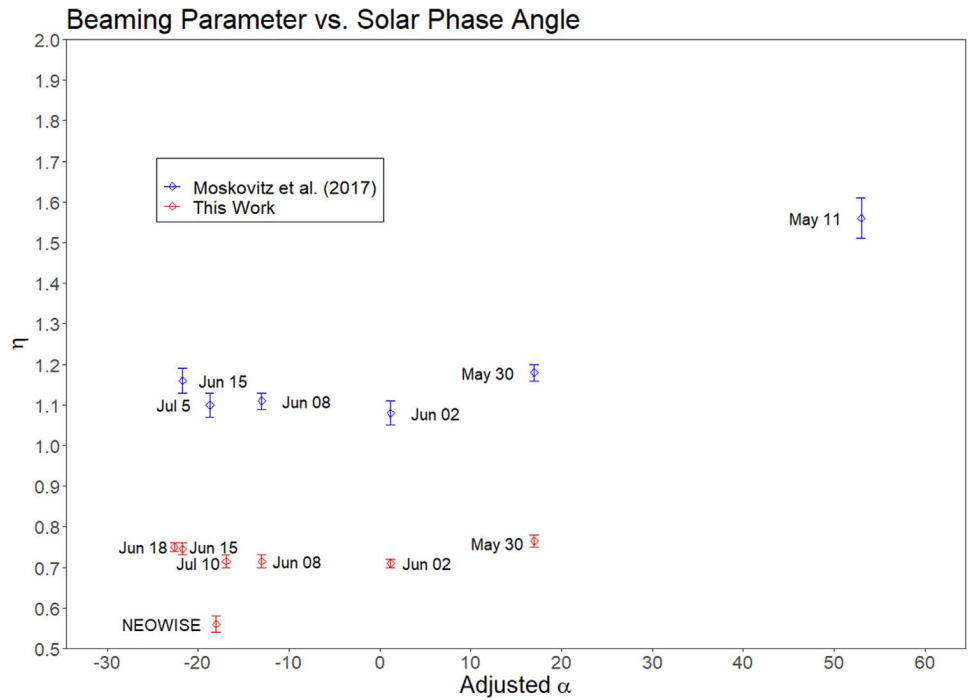

**Figure 8.** Plot of fitted beaming parameters as a function of solar phase angle adjusted so that 0° corresponds to QE2's minimum phase angle during its close approach to Earth. We also compare our beaming parameters to those found by Moskovitz et al. (2017). Note the introduction of negative phase angles to differentiate between observations taken before (positive values) and after (negative values) opposition. The error bars represent the range of beaming parameters. The range is calculated by identifying models that fit the data with fixed albedo and thermal inertia (Section 4). Moskovitz et al. (2017) values are taken from their Figure 3. We see that our data exhibit roughly the same trend where the data overlap, but that our beaming values are significantly offset from the Moskovitz et al. (2017) values.

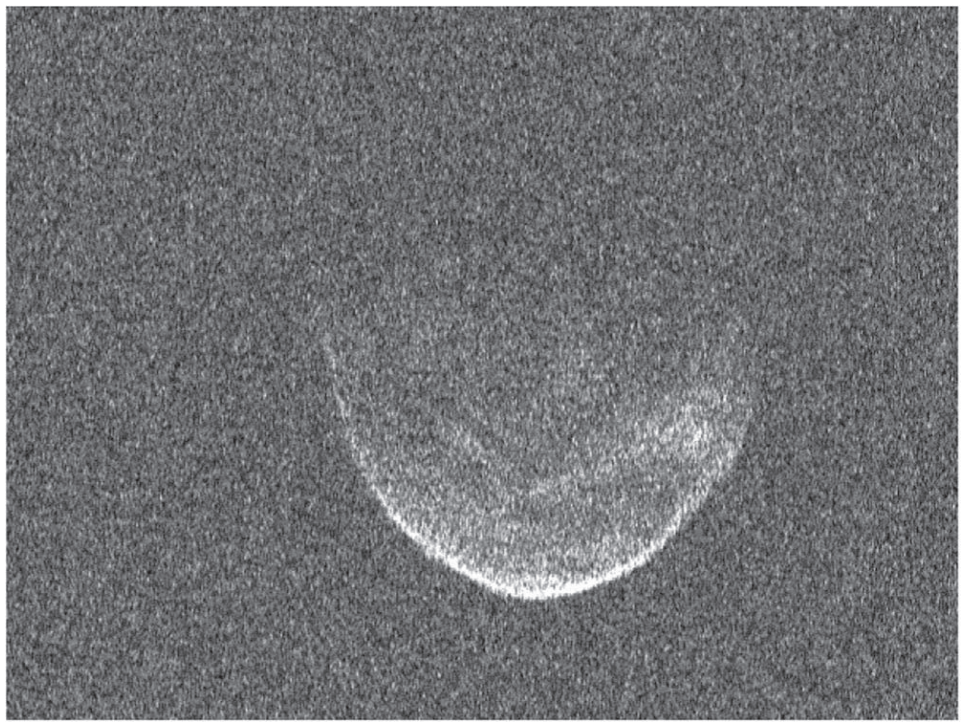

**Figure 9.** Radar image of QE2 taken by the Arecibo Telescope on 2013 June 10. The vertical extent of the image shows distance from the observer to the terminator of the object. The horizontal extent shows Doppler shift, with blueshift to redshift going left to right. The resolution of the pixels, combined with knowledge of the speed of light, directly gives the object's radius. In this image, QE2 covers 210 pixels in the vertical extent at 7.5 m pixel$^{-1}$, giving an apparent radius of 1575 m or a diameter of 3.15 km.

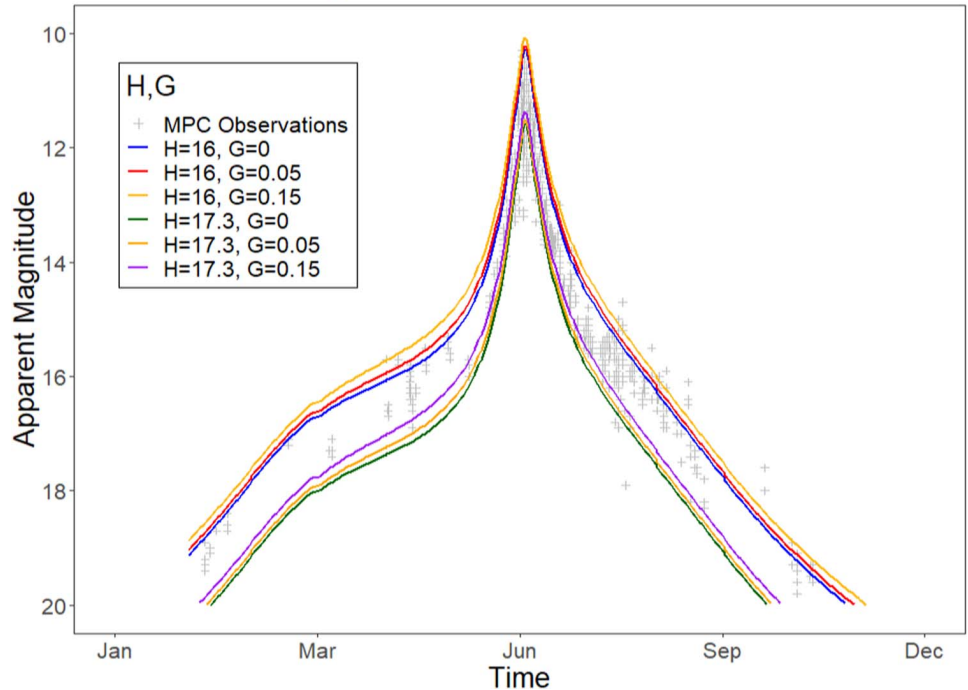

**Figure 10.** Plot of predicted apparent magnitudes for QE2 compared to all magnitudes reported to the MPC. All observations are from 2013 during QE2's close approach to Earth. The predicted apparent magnitudes were calculated using ephemeris from JPL Horizons at 1-day intervals throughout 2013. We used $H = 16.0$ (a value from our modeled $H$-magnitude range) and $H = 17.3$ (the $H$-magnitude from Moskovitz et al. 2017), as well as a range of $G$ values. We see that a lower $H$-magnitude, more consistent with our modeled range, agrees with the data for low $G$ values.

SHERMAN is a forward model, so we generate many models across different values of the free-floating parameters to match to our data. Some preliminary model results are shown in Figure 12. We find that an albedo of 0.053, thermal inertia of 200 TIU, and crater fraction of 70% can roughly match the data. These values are also consistent with the results of the NEATM-like model.

The SHERMAN results also show that the topography of QE2 is affecting the thermal emission. Using SHERMAN, we run models using both possible pole solutions. The results show slight differences in the model fits to the data between these solutions, with a clear preference for the B solution, implying that these features are most likely located in QE2's southern hemisphere (Figure 12). Thus, topography is likely playing a role for QE2 and is likely affecting the uncertainties in the simple thermal model results. Furthermore, topography may be one of the effects being captured by variations in our NEATM-like model's beaming parameter.

### 5.4. Beaming Parameter Trends

The NEATM-like model's beaming parameter is a scaling factor that accounts for additional effects not included in the model. As such, we can analyze the trend in our measured beaming parameters across each night of observation to understand the limitations of our NEATM-like model. We find beaming parameters that range from 0.54 to 0.78. These values therefore differ significantly from the value of $\eta = 1.2$ predicted by Harris (1998) for NEAs. Our modeled beaming parameters are instead much closer to the $\eta \approx 0.75$ value predicted by Lebofsky et al. (1986) for main belt objects. Since the beaming parameter accounts for additional factors not incorporated into the NEATM-like model, we can use these differences to identify potential properties affecting QE2's thermal emission. QE2 is a particularly good target for this analysis owing to its extremely spherical shape. Therefore, shape effects are likely a very small contributor to changes in the beaming parameter.

One potential method for investigating beaming parameters is by looking for trends as a function of solar phase angle. Moskovitz et al. (2017) previously applied this method to QE2. Using beaming parameter as a proxy for thermal emission, Moskovitz et al. (2017) identified QE2 as a prograde rotator. We investigate this trend by showing the phase angle for QE2, which has a minimum value of 17°.1 on June 3, along with the fitted beaming parameters for the best-fit NEATM models for each night. We compare our results to those found by Moskovitz et al. (2017) in Figure 8.

We find that our beaming parameter values do exhibit roughly the same trend as the Moskovitz et al. (2017) data but are significantly offset from the Moskovitz et al. (2017) data. We find much lower beaming parameter values than the Moskovitz et al. (2017) values of $\sim$1.1–1.4. We also find a range of thermal inertias that is overlapping with, but lower than their estimated range of $\sim$200–400 TIU done by comparing their NEATM results to more complex models. This is not unexpected, as our beaming parameter has been separated from the thermal inertia. The Moskovitz et al. (2017) beaming parameter must account for all the effects of thermal inertia, as they do not model thermal inertia explicitly, unlike our NEATM-like model, which does incorporate thermal inertia.

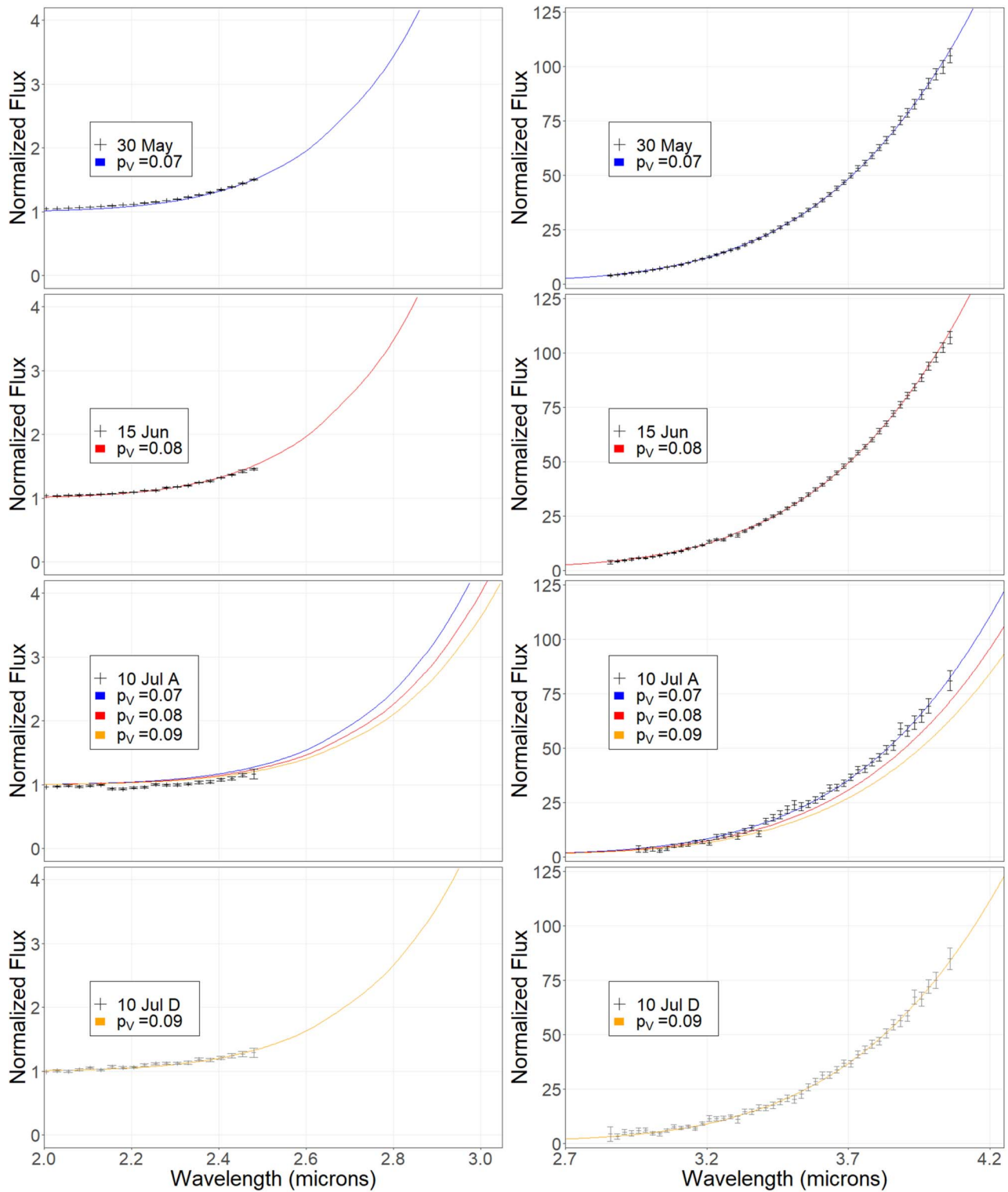

**Figure 11.** Plot of NEATM-like models with varying visual geometric albedos across a selected range of dates. The data sets shown for May 30 and June 15 are the "A" data sets. All models shown have thermal inertia and beaming parameters that are within the best-fit ranges for the given date. Each row is a different data set. The left panels show the tail region, and the right panels show the thermally dominated region. We see that for July 10 A the models that fit the thermally dominated region do not fit the tail region and vice versa. An increase in albedo of ∼0.02 is required to fit the tail region for July 10 A. This is indicative of some kind of surface inhomogeneity.

Another possible explanation for why we observe different beaming parameters is because of our expanded wavelength range (Section 5.2). We incorporate data up to 4.05 μm in our NEATM-like model, while Moskovitz et al. (2017) only incorporate data up to 2.5 μm. As shown in Figure 11, mismatches in model fits between the thermally dominated region and tail region of the spectra are possible. We check this by comparing the Moskovitz et al. (2017) fits to our data at longer wavelengths (Figure 13). As expected, we see that although the Moskovitz et al. (2017) models fit the tail region, they do not fit the thermally dominated region.

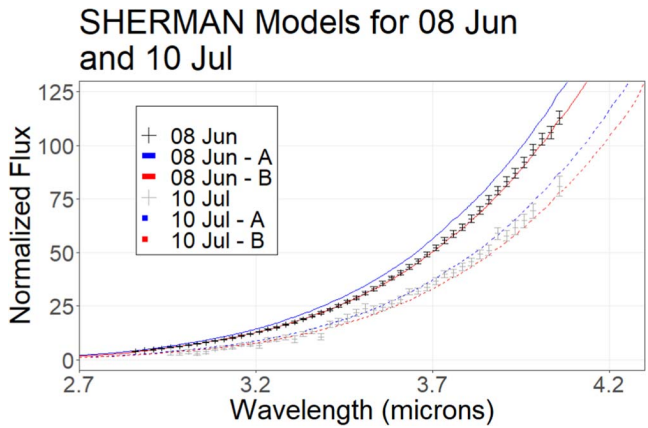

**Figure 12.** SHERMAN model results for June 8 and July 10 using both the A and B pole solutions. All models have a visual geometric albedo of 0.053, thermal inertia of 200 TIU, and crater fraction of 70%. We see a clear preference for the B pole solution in the June 8 data and a slight preference for the B pole solution in the July 10 data. Thus, we see that QE2's topography may be playing a role in shaping its thermal emission. We also note that the albedo and thermal inertia are consistent with our NEATM-like model.

The differences in measured beaming parameters could also be related to the illumination geometry of QE2. The technique used by Moskovitz et al. (2017) relies on assuming that the observations of the asteroids were made with equatorial illumination and thus may not be as robust when viewing an object with a different illumination geometry. (Moskovitz et al. 2017 also recognize this possibility.) Although the observations of QE2 are made at low sub-Earth latitudes, it is possible that the discrepancy in the beaming parameters could arise from high subsolar latitudes. For QE2 these can range from $\sim 30°$ to $\sim 45°$ for the A pole solution or from $\sim 10°$ to $\sim 15°$ for the B pole solution. Thus, because QE2 is not being observed looking directly at its equator, this means that self-shadowing from topographical features on the asteroid's surface is likely to be important. Even for the more equatorial illuminated B pole solution, self-shadowing could still be playing a significant role, as QE2 does not have an equatorial ridge and thus still has topographical variation at the equator. This agrees with our SHERMAN results that show the importance of topography on QE2, which may be contributing to observed temperature differences (Section 5.3). Thus, this may further explain why our beaming parameter results differ from those of Moskovitz et al. (2017).

## 6. Summary and Conclusions

We present simple thermal model fits using our NEATM-like model for the NEA (285263) 1998 QE2. Furthermore, we compare these model results to more complex thermophysical models, radar data, and other existing analyses of QE2 to understand the key factors affecting the uncertainties in simple thermal model results. For our simple thermal model fits, QE2 was observed with the SpeX instrument on the NASA IRTF on six nights in 2013, representing a range of viewing and illumination geometries. Additional data were acquired by the NEOWISE spacecraft in 2017. A visual geometric albedo between 0.05 and 0.10 and thermal inertia between 0 and 425 TIU are found to be consistent with all six nights of SpeX data. These results are also consistent with the NEOWISE absolute photometry at the $3\sigma$ level. These constraints are more robust than they would be using NEOWISE observations alone, due to the larger uncertainties on absolute photometry. The general model agreement with both absolute flux and normalized flux measurements increases our confidence in our model results, while also allowing us to benefit from the smaller uncertainties on normalized flux data. This is possible because of our incorporation of data representing a range of viewing geometries. As a result, we are able to break degeneracies in model results based on a single night of observations.

In order to constrain the limits of simple thermal models as applied to a single object, we compare our results to more complex thermophysical models and previous observations. We find that our modeled albedo values are higher than but overlap with previously published values (Moskovitz et al. 2017; Fieber-Beyer et al. 2020) and are consistent with results from the complex thermophysical model SHERMAN. We also identify a discrepancy in the resulting $H$-magnitude value when using the radar-derived size measurement (Springmann et al. 2014). Based on the tight constraints we place on QE2's albedo and the tighter constraints Springmann et al. (2014) place on QE2's diameter, we believe that the true $H$-magnitude value must be brighter than current measurements suggest. As a result, the true albedo is likely toward the lower end of the range we identify using our NEATM-like model.

We also leverage the wide wavelength range of our data set to compare our best-fit model results to both the tail region and thermally dominated region of our spectra. We find that for some dates, although our models fit the thermally dominated region well, they require a higher albedo to fit the tail region. This highlights the need to incorporate data across a wide wavelength range when modeling asteroid surface properties. We posit that these differences may be due to local thermal variations, but a full investigation is beyond the scope of this work.

In addition to these discrepancies, we also find differences between our modeled beaming parameters and existing models. The most likely source of these differences may be the orientation of QE2 and wavelength range of data used. Observing these differences has also allowed us to infer that topography may play a significant role in determining the thermal emission of QE2. Thus, in this case, the inability to model self-shadowing effects from topographical variations may be a key limiting aspect of the simple thermal models. Furthermore, this analysis again shows the importance of incorporating data from a wide wavelength range when working with simple thermal models.

Overall, our work has demonstrated our ability to place tighter constraints on the results of simple thermal models by comparing data taken across multiple different viewing geometries. By combining normalized flux with absolute photometry, we are able to place tighter constraints than would be possible with absolute photometry alone. Finally, we are able to place some constraints on the limits of simple thermal models as applied to single objects, finding that topography, viewing geometry, and the wavelength range of data used can all affect simple thermal model results. This work is important for diagnosing cases (such as QE2) where more detailed analysis of an object may be required to fully understand its properties.

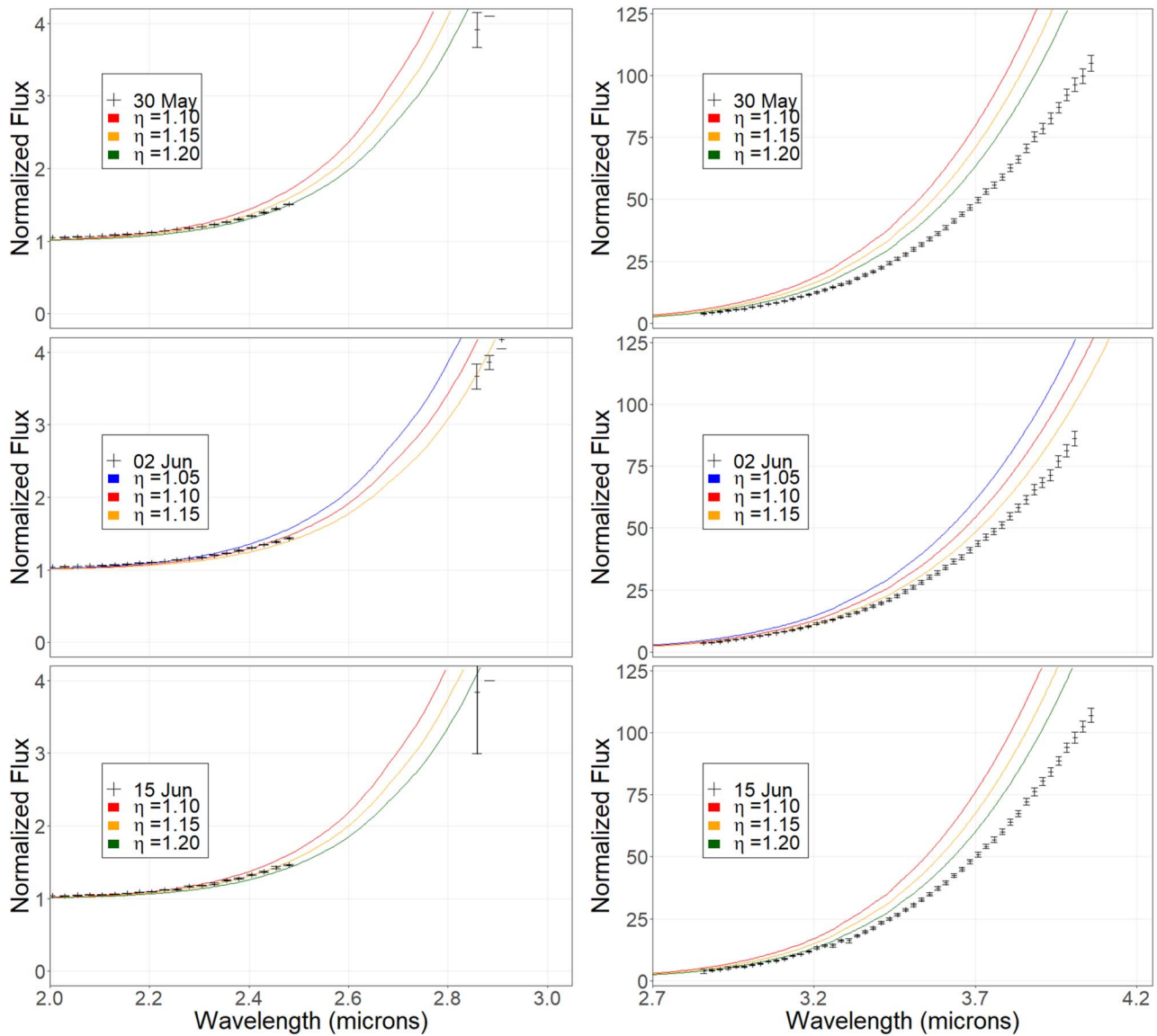

**Figure 13.** Plot of best-fit models from Moskovitz et al. (2017) compared to our longer-wavelength data. These models are generated using our simple, NEATM-like model. All data sets shown are the "A" data set for the given date. All models shown have zero thermal inertia and albedos of 0.03, as per the Moskovitz et al. (2017) fits. The shown $\eta$ values correspond to the ranges reported for each date by Moskovitz et al. (2017). The left panels show the tail region, and the right panels show the thermally dominated region. We see that the models fit the tail region well, as expected. However, we note that these models do not fit the thermally dominated region. This discrepancy may explain why we find different modeled beaming parameters than Moskovitz et al. (2017).

Being able to extract more information from simple thermal models, like our NEATM-like model, will be critical as we move into the future of large survey missions such as LSST and NEO Surveyor. The large data volumes produced by these missions will necessitate the use of simple models to make full use of the data. Using these data as efficiently as possible will require further insights into the limitations of simple thermal models. As this work shows, although these models are reliable for statistical measurements of large groups of objects, the results for individual objects may be subject to great uncertainties. Addressing these issues will therefore allow us to make full use of these models and gain even greater insights into fields such as planet formation, asteroid dynamics, and planetary defense.

This work was partially funded by the NASA YORPD program (NASA grant 80NSSC21K0658) and NSF AST 1856411. S.A.M. was supported by the University of Arizona, Lunar and Planetary Laboratory, Lieutenant Colonel Kenneth Rondo Carson and Virginia Bryan Carson Graduate Fellowship. This material is based on work supported by the National Science Foundation Graduate Research Fellowship Program under grant No. DGE-2137419. Any opinions, findings, and conclusions or recommendations expressed in this material are those of the author(s) and do not necessarily reflect the views of the National Science Foundation. S.E.M. was supported by NASA's Near-Earth Object Observations Program through grant 80NSSC19K0523.

### ORCID iDs

Samuel A. Myers ⓘ https://orcid.org/0000-0001-8500-6601
Ellen S. Howell ⓘ https://orcid.org/0000-0002-7683-5843

Christopher Magri https://orcid.org/0000-0002-2200-4622
Ronald J. Vervack, Jr. https://orcid.org/0000-0002-8227-9564
Yanga R. Fernández https://orcid.org/0000-0003-1156-9721
Sean E. Marshall https://orcid.org/0000-0002-8144-7570
Patrick A. Taylor https://orcid.org/0000-0002-2493-943X

## References

Arai, T., Yoshida, F., Hong, P., et al. 2020, LPSC, 51, 2924
Belton, M., Veverka, J., Thomas, P., et al. 1992, Sci, 257, 1647
Belton, M. J., Chapman, C. R., Klaasen, K. P., et al. 1996, Icar, 120, 1
Bowell, E., Hapke, B., Domingue, D., et al. 1989, in Asteroids II, ed. R. P. Binzel, T. Gehrels, & M. S. Matthews (Tucson, AZ: Univ. Arizona Press), 524
Cushing, M. C., Vacca, W. D., & Rayner, J. T. 2004, PASP, 116, 362
Cutri, R., Wright, E., Conrow, T., et al. 2012, Explanatory Supplement to the WISE All-Sky Data Release Products, NASA/JPL
Delbó, M., Harris, A. W., Binzel, R. P., Pravec, P., & Davies, J. K. 2003, Icar, 166, 116
Dollfus, A. 1971, IAU Colloq. Proc. 12, Physical Studies of Minor Planets (Cambridge: Cambridge Univ. Press), 25
Ďurech, J., Vokrouhlický, D., Baransky, A., et al. 2012, A&A, 547, A10
Fieber-Beyer, S. K., Kareta, T., Reddy, V., & Gaffey, M. J. 2020, Icar, 347, 113807
Harris, A. W. 1998, Icar, 131, 291
Hicks, M., Lawrence, K., Chesley, S., et al. 2013, ATel, 5132, 1
Hinkle, M., Howell, E., Fernández, Y., et al. 2022, Icar, 382, 114939
Howell, E., Magri, C., Vervack, R., Jr., et al. 2018, Icar, 303, 220
Howell, E., Nolan, M., Lejoly, C., et al. 2020, AAS/DPS Meeting, 52, 409.02
Howell, E. S., Vervack, R., Jr., Nolan, M., et al. 2012, AAS/DPS Meeting, 44, 110.07
Hudson, R. S., & Ostro, S. J. 1994, Sci, 263, 940
Jones, J. 2018, PhD thesis, Univ. Central Florida
Jurić, M., Ivezić, Ž., Lupton, R. H., et al. 2002, AJ, 124, 1776
Lagerros, J. S. V. 1998, A&A, 332, 1123
Lauretta, D., DellaGiustina, D., Bennett, C., et al. 2019, Natur, 568, 55
Lebofsky, L. A., & Spencer, J. R. 1989, in Asteroids II, ed. R. P. Binzel, T. Gehrels, & M. S. Matthews (Tucson, AZ: Univ. Arizona Press), 128
Lebofsky, L. A., Sykes, M. V., Tedesco, E. F., et al. 1986, Icar, 68, 239
Magri, C., Howell, E. S., Nolan, M. C., et al. 2011, Icar, 214, 210
Magri, C., Howell, E. S., Vervack, R. J., Jr., et al. 2018, Icar, 303, 203
Magri, C., Ostro, S. J., Scheeres, D. J., et al. 2007, Icar, 186, 152
Mainzer, A., Bauer, J., Cutri, R., et al. 2014, ApJ, 792, 30
Mainzer, A., Bauer, J., Grav, T., et al. 2011a, ApJ, 731, 53
Mainzer, A., Grav, T., Bauer, J., et al. 2011b, ApJ, 743, 156
Marchis, F., Kaasalainen, M., Hom, E. F. Y., et al. 2006, Icar, 185, 39
Marchis, F., & Vega, D. 2014, AGUFM, 2014, P43F–08
Marshall, S. E., Howell, E. S., Magri, C., et al. 2017, Icar, 292, 22
Masiero, J. R., Mainzer, A., Bauer, J., et al. 2021, PSJ, 2, 162
Masiero, J. R., Wright, E., & Mainzer, A. 2019, AJ, 158, 97
Millis, R. L., & Dunham, D. W. 1989, in Asteroids II, ed. R. P. Binzel, T. Gehrels, & M. S. Matthews (Tucson, AZ: Univ. Arizona Press), 148
Morrison, D., & Teller, E. 1995, in Hazards Due to Comets and Asteroids, ed. T. Gehrels, M. S. Matthews, & A. Schumann (Tucson, AZ: Univ. Arizona Press), 1135
Moskovitz, N. A., Polishook, D., DeMeo, F. E., et al. 2017, Icar, 284, 97
Nolan, M. C., Magri, C., Howell, E. S., et al. 2013, Icar, 226, 629
Ostro, S. J. 1985, PASP, 97, 877
Pravec, P., & Harris, A. W. 2007, Icar, 190, 250
Rayner, J., Toomey, D., Onaka, P., et al. 2003, PASP, 115, 362
Springmann, A., Taylor, P., Howell, E., et al. 2014, LPSC, 45, 1313
Taylor, P. A., Howell, E., Nolan, M., et al. 2014, in Asteroids, Comets, Meteors, ed. K. Muinonen et al. (Helsinki) 524
Taylor, P. A., Rivera-Valentín, E., & Aponte-Hernandez, B. 2019, EPSC-DPS Joint Meeting, 51, 689
Trilling, D. E., Mueller, M., Hora, J., et al. 2010, AJ, 140, 770
Vereš, P., Jedicke, R., Fitzsimmons, A., et al. 2015, Icar, 261, 34
Veverka, J., Robinson, M., Thomas, P., et al. 2000, Sci, 289, 2088
Wright, E. L., Eisenhardt, P. R., Mainzer, A. K., et al. 2010, AJ, 140, 1868